\documentclass{IEEEtran}
\usepackage{cite}
\usepackage{amsmath,amssymb,amsfonts}
\usepackage{graphicx}
\usepackage{textcomp,nicefrac}
\usepackage{lipsum}
\usepackage{siunitx}
\usepackage{float}
\usepackage{multirow}

\def\BibTeX{{\rm B\kern-.05em{\sc i\kern-.025em b}\kern-.08em
T\kern-.1667em\lower.7ex\hbox{E}\kern-.125emX}}
\begin{document}
\title{Time-of-Flight X-ray Measurements for Computed Tomography}
\author{\small{G. Lemaire, R. Espagnet, J. Rossignol, L.-D. Gaulin, E. Villemure, A. Corbeil Therrien, Y. Bérubé-Lauzière, M.-A. Tétrault and R.~Fontaine$^{\dagger}$}

\thanks{This work did not involve human subjects or animals in its research.}
\thanks{$^{\dagger}$ Institut Interdisciplinaire d'Innovation Technologique (3IT) and Département de Génie Électrique et Informatique, Université de Sherbrooke, Sherbrooke, Québec, Canada.}
}

\maketitle

\begin{abstract}
    Time-of-flight (ToF) measurements is a possible alternative to anti-scatter grids in computed tomography (CT). Simulations have shown a possible 75\% reduction in the detrimental scattering contribution with a 100~ps FWHM timing resolution. 
    A test bench comprising a pulsed X-ray source and a time-correlated detector has been designed to demonstrate the feasibility of time-of-flight measurements of X-rays using readout electronics inherited from a ToF-positron emission tomography project. A 86~ps FWHM coincidence time resolution have been achieved with 511~keV annihilation gamma-rays and a 155~ps FWHM timing jitter with a 120~kVp pulsed X-rays source. 
\end{abstract}

\section{Introduction}
\label{sec:intro}

    Computed tomography (CT) plays a key role in medical imaging, thanks to its high spatial resolution for volumetric images, its low cost, and short acquisition time compared to other modalities such as magnetic resonance imaging. CT imaging uses ionizing radiation, and therefore delivers a certain dose to the patient for each acquired image. The deposited dose is directly linked to the probability of radiation-induced cancer, either by causing non-repairable DNA damage, or by increasing the probability of long-term stochastic effects~\cite{cancer}. It should therefore be kept as low as possible. 


    One of the main factor in image degradation in CT is the contribution of scattered photons. The image is obtained by measuring the transmission of an X-ray beam as it passes through the patient, with attenuation being indicative of the nature of tissues. The useful information is provided by the photons that have travelled in a straight line, as they did not interact with the material they went through. Scattered photons, whose trajectories have been deviated from a straight line, can nevertheless be detected in adjacent pixels and thus contribute to the noise in the image, as they do not satisfy the hypothesis of straight line propagation, as assumed by image reconstruction. This noise degrades the contrast-to-noise ratio, while distorting the absorption measurement.

    Scattered photons can be mechanically discarded by using an antiscatter grid, or statistically removed in post-processing. The former approach reduces the sensitivity, and thus requires a higher dose and the later is subject dependant~\cite{scatter}. 

    Time-of-flight (ToF) measurements have recently been propose as a solution to avoid the aforementioned problems~\cite{julien}. ToF-CT, also known as Time-Correlated Single Photon Counting (TCSPC)~\cite{tcspc}, relies on measuring the time-of-flight of X-rays between a pulsed X-ray source and time-correlated detectors. The approach enables discriminating between ballistic photons travelling along straight  paths, and thus having shorter travel times, compared to scattered photons. Owing to the 10~ps ToF quest initiated in Positron Emission Tomography (PET)~\cite{lecoq}, improvement in scintillators and photodetectors have been made and measurement of ToF of X-rays is now within reach.

    The temporal resolution of the detection system is therefore directly linked to the reduction in the detrimental contribution of scattered photons to image noise. Simulations carried out with \textsc{GATE}~\cite{gate} have shown that a system with a temporal resolution of 200~ps FWHM can achieve a 66\% reduction in the scattered to primary ratio (SPR). This reduction reaches 75\% at 100~ps FWHM and 95\% at 10~ps FWHM~\cite{julien}. In comparison, commercially available anti-scatter grids achieve reductions in the order of 70 to 95\% of the contribution of scattered photons, depending on grid characteristics~\cite{antiscat2}. However, this reduction is inevitably accompanied by a reduction in primary fluence, of 15 to 30\%, and therefore by an increase in patient dose to compensate for the grid-induced fluence loss~\cite{antiscat1}.

    This paper presents the first measurements made on a tabletop prototype dedicated to demonstrate the feasibility of ToF-CT. This prototype is a part of the on-going ToF-CT project at 3IT (Université de Sherbrooke, QC, Canada). This paper first presents the materials used in the experiments, followed by the data acquisition and analysis methods, and then by the main results.
    
\section{Materials}
\label{sec:material}

    All measurements were carried out on an SDA 6000A \textsc{LeCroy} (Chestnut Ridge, NY, USA) oscilloscope. The acquired data were processed using Python. The test bench comprises a pulsed X-ray source and a time-correlated detector, described in detail in the following subsections.

    \subsection{Laser-Triggered Pulsed X-Ray Source}

        The X-ray source was manufactured by \textsc{Hamamatsu} (Hamamatsu City, Japan). It comprises a pulsed laser and an \mbox{X-ray} tube. The laser, model PLP10-040 with control module C10196, emits 60~ps FWHM pulses at 405~nm, at a frequency adjustable up to 100~MHz. The laser control module features a synchronization channel with a time interval error jitter (TIE) below 10~ps when working at 1~MHz. The laser excites a photocathode inside the X-ray tube to create bunches of electrons. These electrons are accelerated towards a tungstene target which then produces the x-rays. The X-ray source accepts average tube currents up to 10~$\mu$A and high voltages up to 120~kVp. A summary of its characteristics is given in Table~\ref{fig:source}.


        \begin{table}[H]
            \centering
            \caption{Characteristics of the X-ray source.}
            \centering
            \includegraphics[scale=0.35]{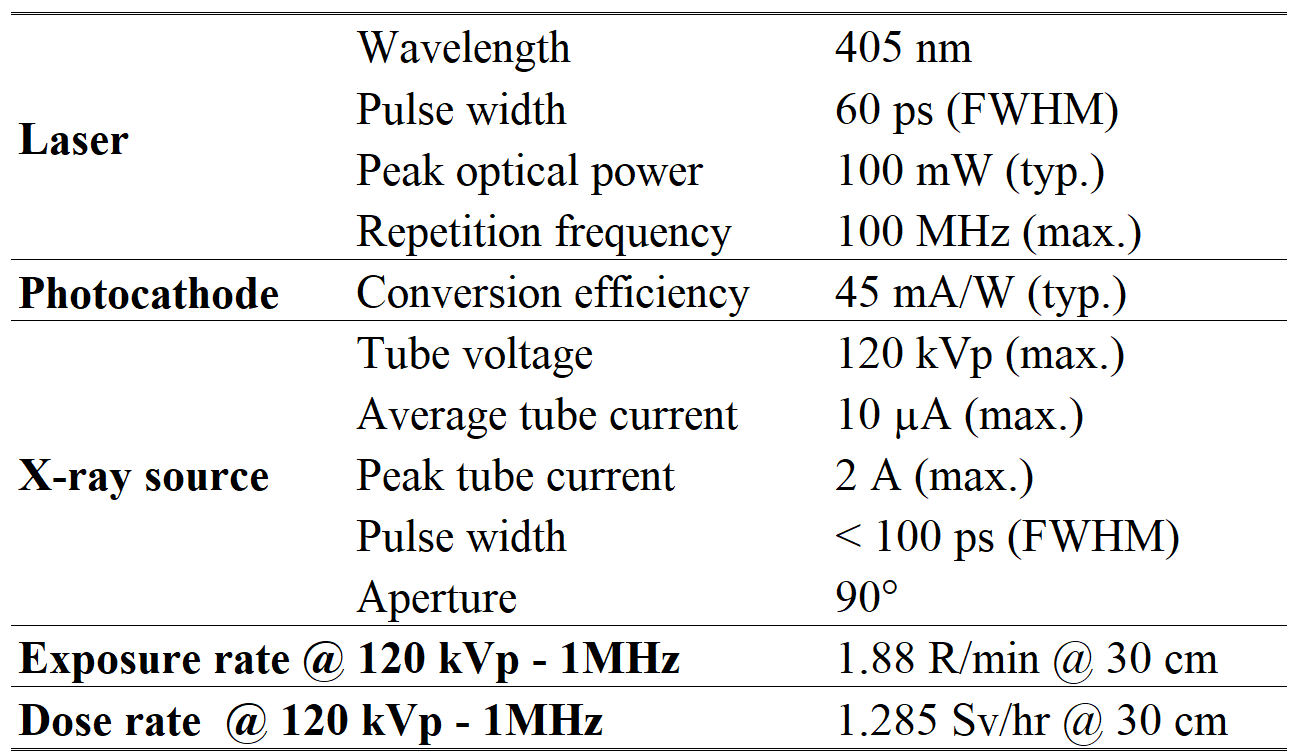}
            \label{fig:source}
        \end{table}

    \subsection{Detector}

        ToF measurements are enabled by a scintillation-based detector using a crystal scintillator, a silicon photomultiplier (SiPM) and low-noise, low-jitter, and high-frequency (HF) readout electronics. Two different scintillators have been tested: a 38.6~ns decay time cerium-doped LYSO from Hilger Crystals and a 45.7~ns decay time cerium-doped LGSO from Hitachi Chemicals both with paint coating. The SiPM used for those measurements is the AFBR-S4N44P014M from \textsc{Broadcom} (San Jose, CA, USA). Scintillators and SiPM charateristics are summarized respectively in Tables~\ref{fig:scintillator} and \ref{fig:SiPM}. 

        \begin{table}[H]
            \centering
            \caption{Characteristics of scintillators.}
            \centering
            \includegraphics[width=\linewidth]{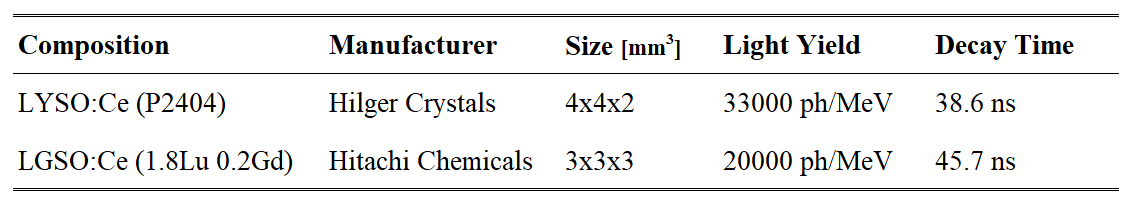}
            \label{fig:scintillator}
        \end{table}

        \begin{table}[H]
            \centering
            \caption{Characteristics of SiPM.}
            \centering
            \includegraphics[width=\linewidth]{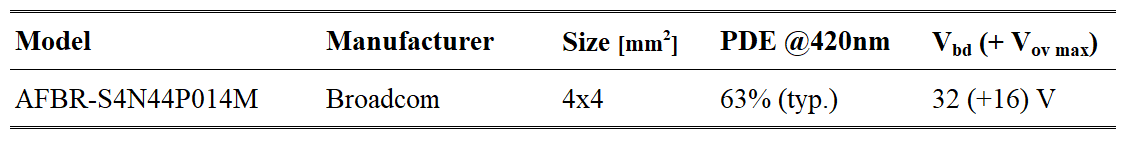}
            \label{fig:SiPM}
        \end{table}
        
\begin{figure}[H]
            \centering
            \includegraphics[width=\linewidth]{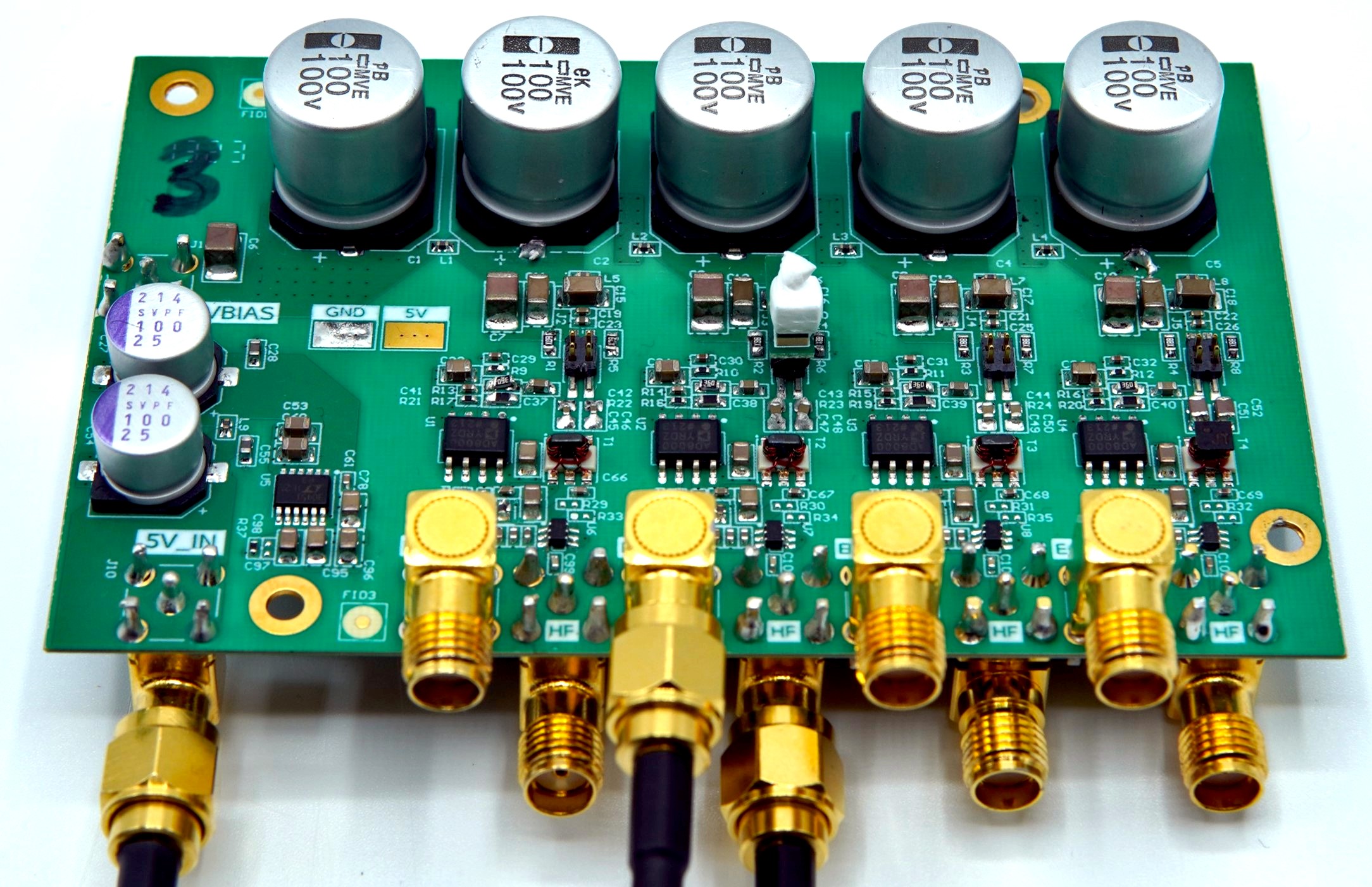}
            \includegraphics[width=\linewidth]{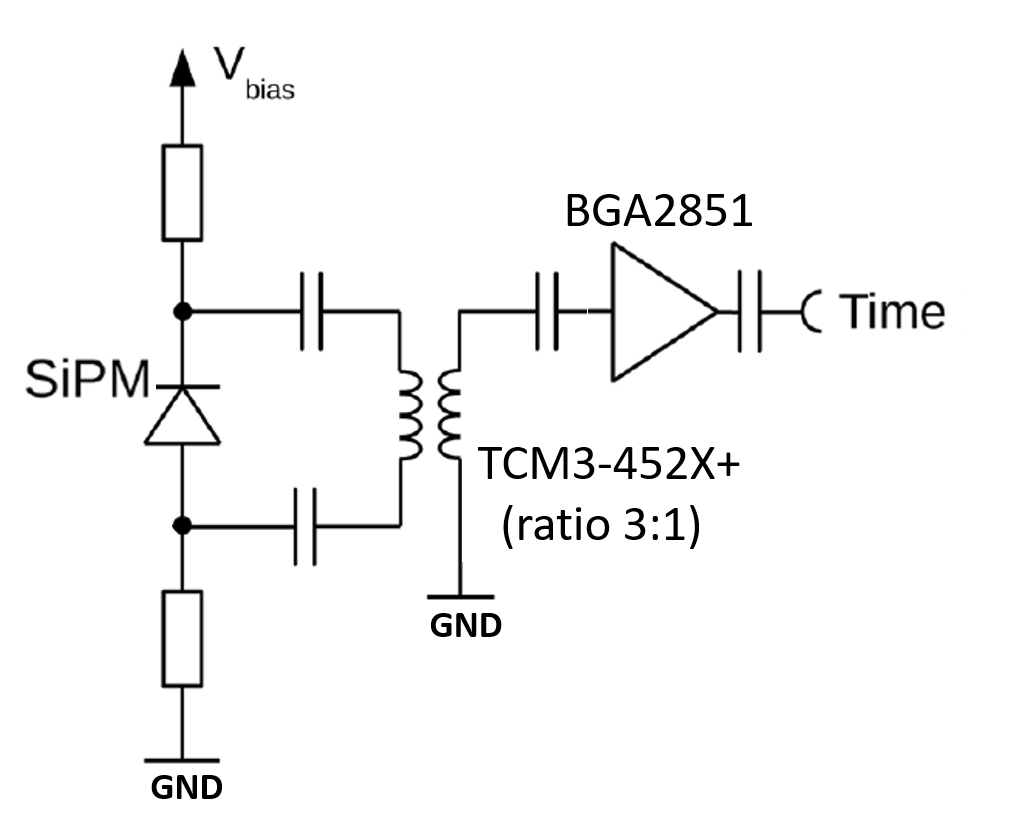}
            \caption{Photo and schematics of SiPM readout electronics, based on ~\cite{catesAMP}.}
            \label{fig:electronics}
        \end{figure}

        The readout electronics, shown in Fig.~\ref{fig:electronics}, is based on an architecture presented in Cates \textit{et al.}~\cite{catesAMP} and Gundacker \textit{et al.}~\cite{gundacAMP}. It comprises a TCM3-452X+ 3:1 ratio balun transformer from \textsc{Mini-Circuits} (Brooklin, NY, USA) for impedance matching and capacitance compensation, and a BGA2851 HF amplifier from \textsc{NXP USA Inc.} (Austin, TX, USA). Its purpose is to amplify the single-photon response from a biased SiPM while ensuring a good signal-to-noise ratio, and thus optimize the system's time response.

\section{Methods}
\label{sec:methods}

    The main goal of the measurements presented herein is to assess the detection system's temporal resolution, or more precisely the detection chain's timing jitter. This variable defines the stability of the system's temporal response, and represents the average variation in the value obtained for each measurement of the same quantity, or in other words the uncertainty.
    
\begin{figure}[H]
        \centering
        \includegraphics[width=\linewidth]{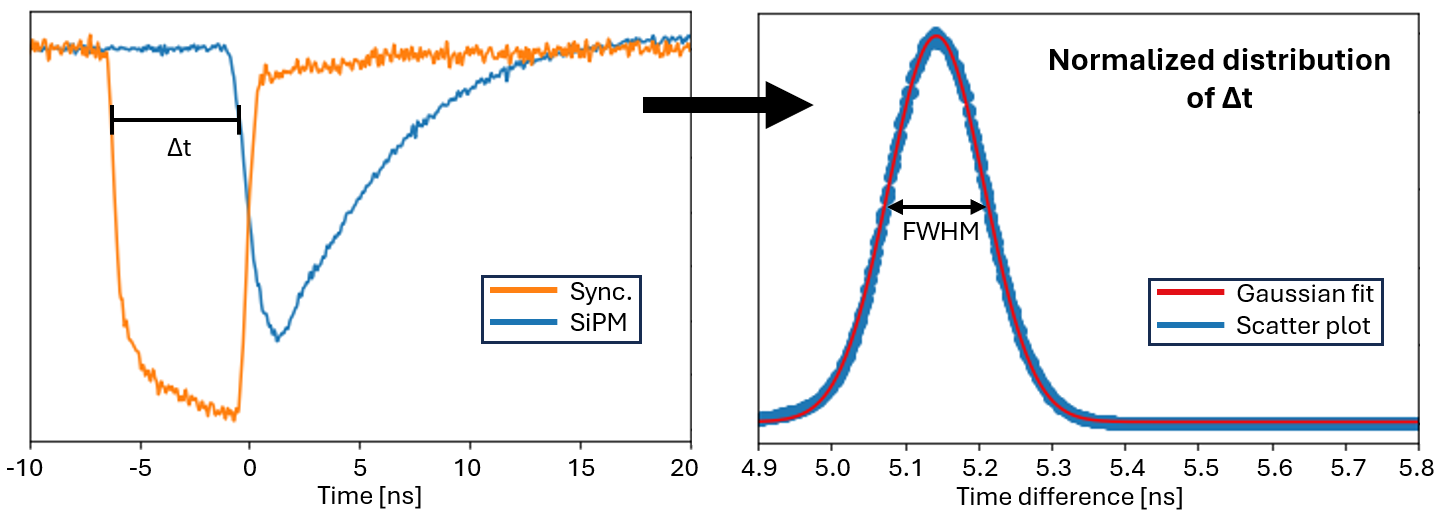}
        \caption{Methodology for data acquisition and analysis.}
        \label{fig:analysis}
    \end{figure}

    The most direct way to measure this uncertainty is to build up a representative sample of time measurements, in order to study their distribution and extract their variance. In the ideal case, the distribution obtained is Gaussian, and the jitter is calculated directly by determining the FWHM of the normal distribution fitted to the data, as shown in Fig.~\ref{fig:analysis}. 

    Two configurations were tested: a positron emission tomography-type configuration with a $^{68}$Ge source, and a tabletop X-ray configuration with the laser-controlled pulsed \mbox{X-ray} source described above and operated at 120 kVp. 

    The typical shape of the measured signals is shown in Fig.~\ref{fig:analysis}. The synchronization signal supplied by the laser control module is shown in orange, while the detector output signal is shown in blue. In both cases, the pulses have a negative amplitude, so the oscilloscope trigger values and thresholds associated with measurements will be negative voltages given in mV. In the next sections, the expression \textit{trigger threshold} will refer to the voltage chosen as the trigger threshold for signal acquisition by the oscilloscope, while the expression \textit{measurement threshold} will refer to the position chosen on the signal waveform to perform the measurement.

    Timing measurements and the histogram of the measurement distribution are performed directly on the oscilloscope. The histogram is then exported to be fitted with a Gaussian function and its FWHM extracted. The oscilloscope is used in logic trigger mode to select time-correlated events, and a trigger threshold is associated with each of the two signals to ensure that they represent true events. The measurement thresholds are then scanned to select the optimized value that minimizes the dispersion of the time measurement distribution.

    No thermal management is performed at the detector level. The dark counts that appear at the start of a pulse can modify the signal baseline and thus distort measurements, as the thresholds do not correspond to the same position on the pulse if it is vertically translated~\cite{darkcount}. This has the effect of complicating the obtained distributions by adding a second Gaussian corresponding to another measurement threshold. Under present measurement conditions, this phenomenon cannot be avoided, and a double Gaussian fit is required to extract the main Gaussian from the measurement distribution.




        

\vspace{0.5cm}

        \begin{figure}[H]
            \centering
            \includegraphics[width=\linewidth]{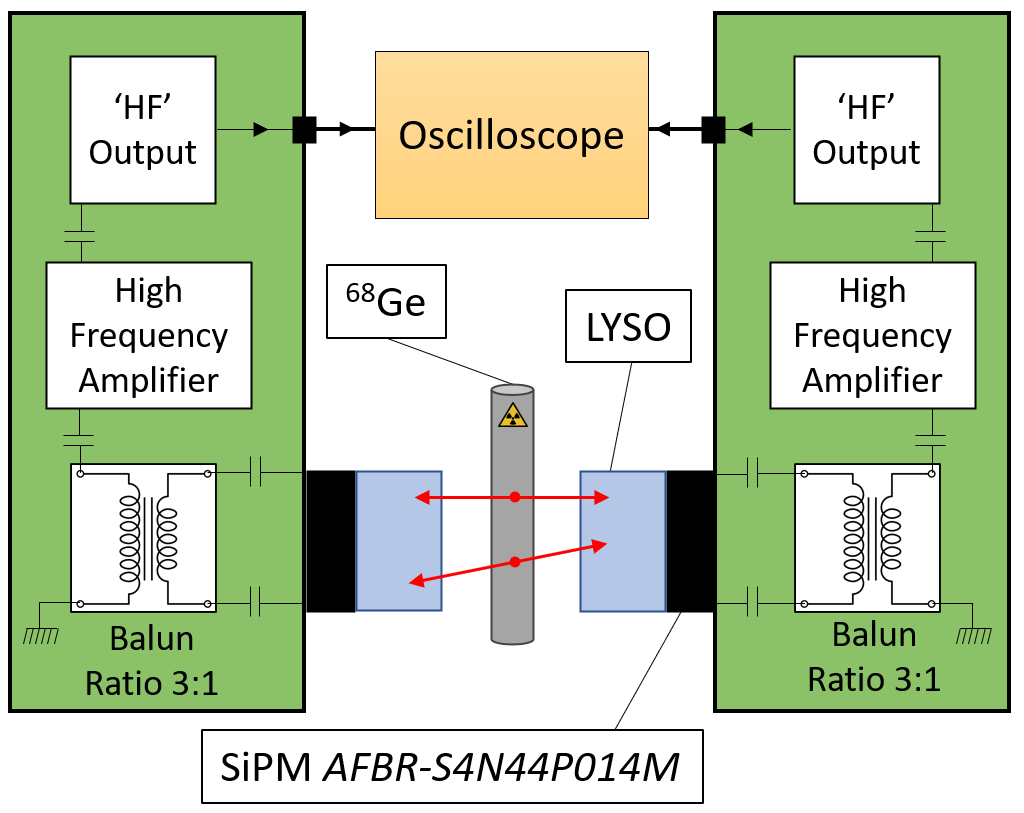}
            \caption{Schematic of the setup used to measure coincidence time resolution (CTR).}
            \label{fig:CTR}
        \end{figure}

\newpage

    \subsection{Coincidence Time Resolution}

        To compare the performance of the analog readout with the literature, a coincidence time resolution (CTR) measurement was performed using a $^{68}$Ge source and two coincident detection chains (Fig.~\ref{fig:CTR}). This measurement was done using the AFBR-S4N44P014M coupled with a LYSO scintillator.

        The SiPM was biased at 40~V, half of the maximum overvoltage value given in the datasheet. The trigger threshold is set at -520~mV for both signals, to avoid events corresponding to energies below 511~keV. The same measurement threshold of -160~mV is used for both signals.


    \subsection{X-Ray Time-of-Flight Jitter}

\begin{figure}[H]
            \centering
            \includegraphics[width=\linewidth]{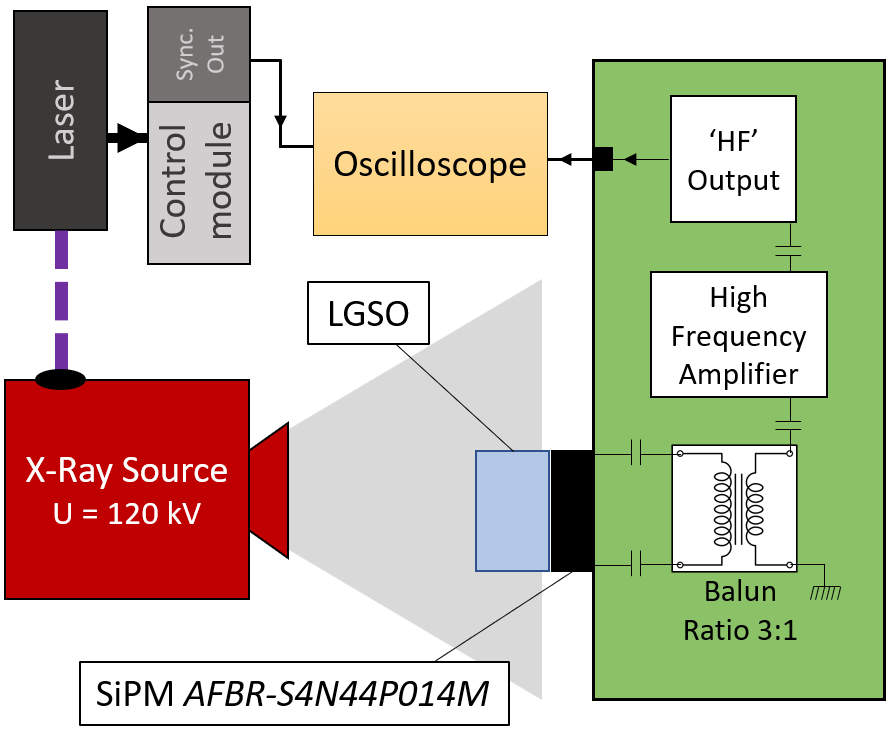}
            \caption{Schematic of the whole test bench for ToF measurements of X-rays.}
            \label{fig:TOF}
        \end{figure}

        The whole system as shown in Fig.~\ref{fig:TOF}, was evaluated. The detector has been irradiated with the pulsed \mbox{X-ray} source operating at 120~kVp and at a 1~MHz repetition rate. The measurement was performed using the AFBR-S4N44P014M and a LGSO scintillator.

        The time measurement is performed between the laser synchronization signal and the detector output signal of the HF amplifier. The SiPM was biased at 40~V. The trigger threshold is set at -485~mV at the detector output, to avoid radiative background noise and low-energy events. The measurement thresholds are swept again, and the value of -50~mV is selected for both signals.

\section{Results}
\label{sec:results}


        
        

    \subsection{Coincidence Time Resolution}

\begin{figure}[H]
            \centering
            \includegraphics[width=\linewidth]{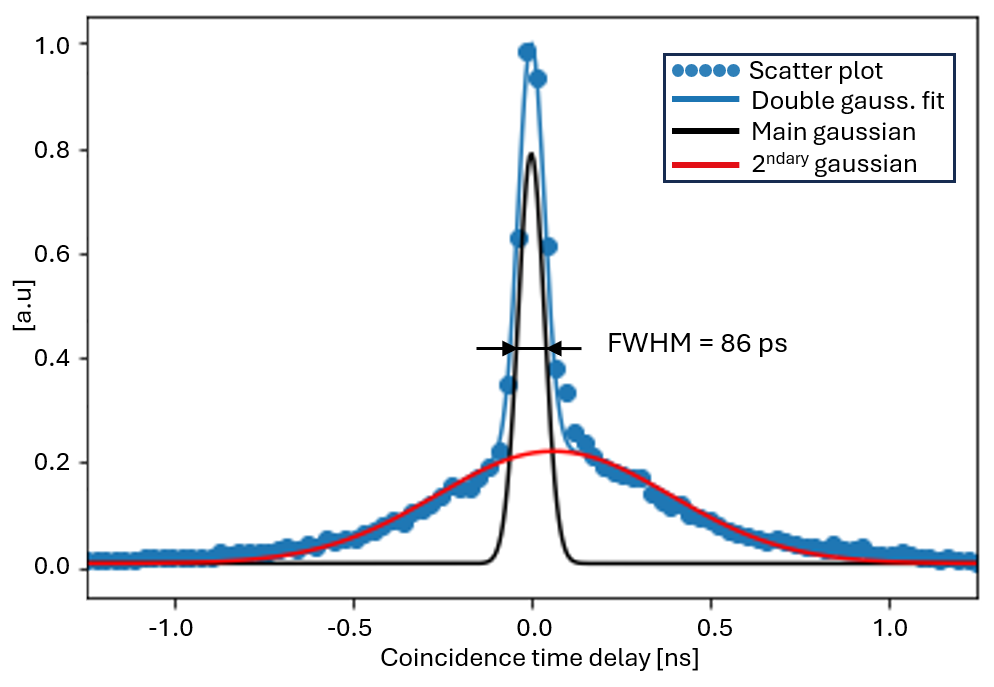}
            \caption{CTR: 511~keV from $^{68}$Ge detected by AFBR-S4N44P014M (Vbias~=~40~V) coupled with LYSO. Data fitted with double Gaussian function.}
            \label{fig:R_CTR}
        \end{figure}

        The fitted data for the coincidence time resolution are shown in Fig.~\ref{fig:R_CTR}. The baseline shift phenomenon is present on both signals. The resulting Gaussian appears wider on the distribution.
        
        A coincidence time resolution of 86~ps FWHM is achieved after extracting the main Gaussian from the double Gaussian fit.
        
    \subsection{X-Ray Time-of-Flight Jitter}

\begin{figure}[H]
            \centering
            \includegraphics[width=\linewidth]{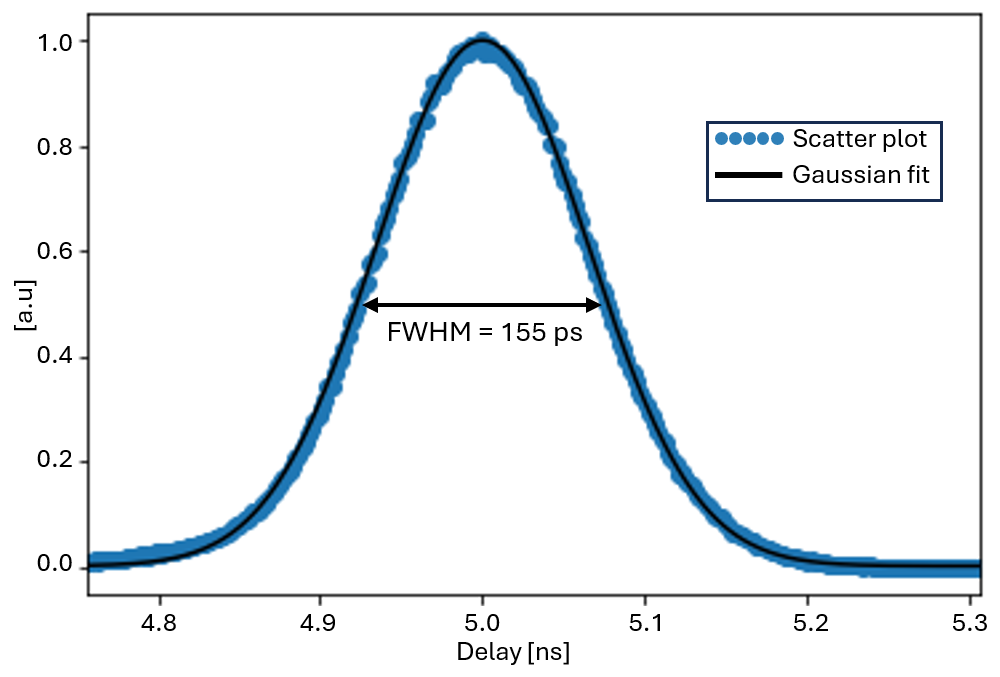}
            \caption{X-Ray ToF jitter: Distribution of delays between X-ray pulse emission (120 kVp, 1 MHz) and detection by AFBR-S4N44P014M (Vbias = 40V) coupled with LGSO. Data fitted with Gaussian function.}
            \label{fig:R_TOF_LGSO}
        \end{figure}

        The fitted data for the time-of-flight distribution is shown in Fig.~\ref{fig:R_TOF_LGSO}. Due to the greater dispersion of measurement results in this configuration, the influence of the baseline shift by dark counts is negligible, and a simple Gaussian fit is sufficient to analyze the resulting distribution.
        
        A time resolution of 155~ps FWHM is achieved with LGSO scintillator.
        
\section{Discussion}
\label{sec:discu}

    The results obtained for coincidence measurements are comparable with the state-of-the-art~\cite{gundacSOTA}, giving confidence in the results obtained with the X-ray source.
    
    With regards to the CTR measurement, as mentionned in the previous sections, the double Gaussian distribution can be explained by the proportion of dark counts during the measurement. Indeed, further measurements have shown a direct link between the bias voltage of the SiPM and the importance of the second Gaussian in the distribution obtained, which disappears at sufficiently low bias voltages but at the cost of timing resolution. Several approaches are being considered to counter this problem: managing the operating temperature, avoiding or reducing parasitic sources, optimising bias voltage.
    

     A complete characterization of the timing jitter of the detection line will require measuring the single photon time resolution (SPTR) of the detector in a photon starving regime. It will then be possible to identify the contributions of the various parts of the system to the total jitter, namely synchronisation signal jitter, X-ray pulse width, SiPM SPTR, noise-induced jitter and setup jitter. The results obtained can then be corrected for these different components.
    
    To have a correct evaluation of the jitter defined above, one would have to precisely measure the time profile of the X-ray pulse. However, no equipment is currently suitable for this type of measurement at the energies considered here, and with the necessary temporal resolution. The time course of the laser pulse used to trigger X-ray generation can be used as a lower limit in the first instance.

    For X-ray measurements, the trigger threshold applied to the detector's output signal does not allow sufficiently accurate energy selection. Future measurements will be carried out with a detector featuring a dedicated channel enabling true energy calibration of the detector.

\section{Conclusion}
\label{sec:conclu}

    This paper presents the first measurements taken on the ToF-CT test bench developed at 3IT, Sherbrooke, including the first time-of-flight measurements of 120~kVp pulsed X-rays. The 86~ps FWHM CTR obtained with SiPM-based detectors is consistent with the state-of-the-art in ToF-PET. Moreover, the temporal resolution of 155~ps FWHM achieved during those tests may enable a reduction of $\sim$~70\% in scatter to primary ratio (SPR), without reducing the primary fluence. Improvements of the system are planned but the results obtained so far meet the expectations.

\section*{Acknowledgement}

    The research team would like to thank Hamamatsu, in particular the Hamamatsu Photonics K.K. subsidiary, for supplying the pulsed X-ray source that makes this research project possible, and Broadcom for supplying the SiPM use for the measurements presented herein. The team also wants to thank Catherine Pépin for her help and the advice given since the beginning of those tests. All authors declare that they have no known conflicts of interest in terms of competing financial interests or personal relationships that could have an influence or are relevant to the work reported in this paper.


\begin{thebibliography}{00}
    \bibitem{cancer}
    E. C. Lin, "Radiation Risk From Medical Imaging", Mayo Clinic Proceedings, vol. 85, no. 12, p. 1142‑1146, dec. 2010, doi: 10.4065/mcp.2010.0260.
    
    \bibitem{scatter}
    E.-P. Rührnschopf et K. Klingenbeck, "A general framework and review of scatter correction methods in x-ray cone-beam computerized tomography. Part 1: Scatter compensation approaches: Scatter compensation approaches", Med. Phys., vol. 38, no. 7, p. 4296‑4311, june 2011, doi: 10.1118/1.3599033.

    \bibitem{julien}
    J. Rossignol et al., "Time-of-flight computed tomography - proof of principle", Phys Med Biol, vol. 65, no. 8, p. 085013, apr. 2020, doi: 10.1088/1361-6560/ab78bf.

    \bibitem{tcspc}
    D. Gaudreault, J. Rossignol, Y. Berube-Lauziere, et R. Fontaine, "Comparative Study of Image Quality in Time-Correlated Single-Photon Counting Computed Tomography", IEEE Trans. Radiat. Plasma Med. Sci., vol. 5, no 3, p. 343‑349, may 2021, doi: 10.1109/TRPMS.2020.3017702.

    \bibitem{lecoq}
    P. Lecoq, "Pushing the Limits in Time-of-Flight PET Imaging", IEEE Trans. Radiat. Plasma Med. Sci., vol. 1, no. 6, p. 473‑485, nov. 2017, doi: 10.1109/TRPMS.2017.2756674.
        
    \bibitem{gate}
    S. Jan et al., "GATE: a simulation toolkit for PET and SPECT", Phys. Med. Biol., vol. 49, no. 19, p. 4543‑4561, oct. 2004, doi: 10.1088/0031-9155/49/19/007.
        
    \bibitem{antiscat2}
    W.-H. Chung, "Study of Monte Carlo Simulator for Estimation of Anti-Scatter Grid Physical Characteristics on IEC 60627:2013-Based", AJPA, vol. 6, no. 2, p. 35, 2018, doi: 10.11648/j.ajpa.20180602.12.
    
    \bibitem{antiscat1}
    J. M. Boone et al., "Development and Monte Carlo Analysis of Antiscatter Grids for Mammography", Technol Cancer Res Treat, vol. 1, no. 6, p. 441‑447, dec. 2002, doi: 10.1177/153303460200100604.

    \bibitem{catesAMP}
    J. W. Cates, S. Gundacker, E. Auffray, P. Lecoq, et C. S. Levin, "Improved single photon time resolution for analog SiPMs with front end readout that reduces influence of electronic noise", Phys. Med. Biol., vol. 63, no. 18, p. 185022, sept. 2018, doi: 10.1088/1361-6560/aadbcd.
    
    \bibitem{gundacAMP}
    S. Gundacker, R. M. Turtos, E. Auffray, M. Paganoni, et P. Lecoq, "High-frequency SiPM readout advances measured coincidence time resolution limits in TOF-PET", Phys Med Biol, vol. 64, no. 5, p. 055012, feb. 2019, doi: 10.1088/1361-6560/aafd52.

    \bibitem{darkcount}
    J. Y. Yeom, R. Vinke, et C. S. Levin, "Optimizing timing performance of silicon photomultiplier-based scintillation detectors", Phys. Med. Biol., vol. 58, no. 4, p. 1207‑1220, feb. 2013, doi: 10.1088/0031-9155/58/4/1207.
        
    \bibitem{gundacSOTA}
    S. Gundacker et al., "Experimental time resolution limits of modern SiPMs and TOF-PET detectors exploring different scintillators and Cherenkov emission", Phys. Med. Biol., vol. 65, no. 2, p. 025001, jan. 2020, doi: 10.1088/1361-6560/ab63b4.
    
\end{thebibliography}
\end{document}